 \documentclass[final,5p,times,twocolumn,authoryear]{elsarticle}

\usepackage{amssymb}
\usepackage{aas_macros}
\usepackage{lipsum}
\begin{document}

\begin{frontmatter}



\title{Glitching pulsars as gravitational wave sources}


\author[first]{B. Haskell}
\affiliation[first]{organization={Nicolaus Copernicus Astronomical Center of the Polish Academy of Sciences},%Department and Organization
            addressline={Bartycka 18}, 
            city={Warsaw},
            postcode={00-716}, 
            country={Poland}}

\author[second]{D.I.~Jones}
\affiliation[second]{organization={Mathematical Sciences and STAG Research Centre, University of Southampton},%Department and Organization
            addressline={}, 
            city={Southampton},
            postcode={SO17 1BJ},
            country={United Kingdom}}
\begin{abstract}

%Mathematical Sciences and STAG Research Centre, University of Southampton, Southampton SO17 1BJ, United Kingdom

Spinning neutron stars, when observed as pulsars, are seen to undergo occasional spin-up events known as glitches.  Despite several decades of study, the physical mechanisms responsible for glitches are still not well understood, but probably involve an interplay between the star's outer elastic crust, and the superfluid and superconducting core that lies within.  Glitches will be accompanied by some level of gravitational wave emission.  In this article, we review proposed models that link gravitational wave emission to glitches, exploring both short duration burst-like emission, and longer-lived signals.  We illustrate how detections (and in some cases, non-detections) of gravitational signals probe both the glitch mechanism, and, by extension, the behaviour of matter at high densities.

\end{abstract}



\begin{keyword}
%% keywords here, in the form: keyword \sep keyword, up to a maximum of 6 keywords
gravitational waves \sep pulsar glitches \sep neutron stars 



\end{keyword}

\end{frontmatter}

%\tableofcontents



\section{Introduction}
\label{introduction}

Pulsar glitches are sudden spin-up events that have now been observed in over a hundred pulsars, and are thought to be linked to the presence of superfluid components in the neutron star interior (for a review see \citealt{AntoReview}). This interpretation is based on the fact that a superfluid rotates by forming an array of quantised vortices, and can only spin-down by expelling some of these vortices. Vortices, however, can be strongly attracted, i.e. `pin', to ions in the crust, or alternatively to superconducting flux tubes in the core of the neutron star. In this case they cannot move out, and the superfluid cannot follow the regular spin-down of the normal component of the star, which is driven by electromagnetic (or possibly gravitational) wave emission. The superfluid thus decouples from the normal component and stores angular momentum, until a large enough lag is built up that hydrodynamical lift forces become strong enough to overcome the pinning forces, leading to sudden vortex motion and angular momentum exchange, i.e. a glitch \citep{1975AndItoh}.

While this qualitative picture was established soon after the first observations of pulsar glitches, the exact details of the mechanism that triggers a glitch, and drives the post-glitch response, or `relaxation' of the frequency of the star, are still debated. Vortex-vortex interactions (or `avalanches'), multifluid instabilities and crust quakes all remain possible explanations for glitch triggers, and while the relaxation is generally thought to be due to the mutual friction coupling between the normal and superfluid components, it is still not clear which regions of the star may be responding, and to what extent crustal moment of inertia re-arrangements following crust quakes may play a role (see \citealt{HMReview} for a review).

Depending on the exact physical mechanisms that are at play, Gravitational Wave (GW) emission may accompany the different phases of a glitch. Bursts of GWs may occur during the glitch itself, due to sudden rearrangements of the superfluid vortices or of the crustal lattice that lead to transient mass or current quadrupoles, while longer lived GW signals may be present during the recovery phase of the glitch, possibly due to modes of oscillation being excited after the glitch. Finally the evolution of the pinned superfluid on long time scales, or permanent changes in the crustal moment of inertia, may lead to persistent signals from glitching pulsars. 

In Section 2 we review the main mechanisms for GW emission that we consider.  In Section 3 we give a brief summary of the spin evolution of a glitching pulsar.  Sections 4, 5 and 6 then describe the GW emission expected during the different glitch phases.  In Section 7 we give a summary of the issues involved in detecting such GW signals.  We finish with a few conclusions in Section 8.

%Citation test \citep{2020MNRAS.498.3138Y}

\section{Gravitational wave emission mechanisms}

GW emission from a system requires, at least to leading order, a time varying mass and (or) current quadrupole moment. The GW amplitude in the transverse traceless gauge can be written as \citep{Thorne:1980ru}
\begin{eqnarray}
h^{TT}_{ij}&=&\frac{G}{c^4 r} \sum_{l=2}^\infty \sum_{m=-l}^l \left( \frac{d^l}{(dt)^l} I_{lm} (t-r) T^{E2}_{lm,ij} \right. \nonumber \\
&&\left. + \frac{d^l}{(dt)^l} S_{lm} (t-r) T^{B2}_{lm,ij}\right) ,
\end{eqnarray}
where $T^{E2}_{lm}$ and $T^{B2}_{lm}$ are electric and magnetic tensor spherical harmonics and $I_{lm}$ and $S_{lm}$ are mass and current multipole moments. 
At the leading order $l=2$, the mass quadrupole $I_{2m}$ and current quadrupole $S_{2m}$ can be expressed as \citep{Thorne:1980ru}
\begin{eqnarray}
I_{2m}&=&\frac{16\sqrt{3}\pi}{15}\int \tau_{00} Y_{2m}^{*}r^2 d^{3}x , \\
S_{2m}&=&\frac{32\sqrt{2}\pi}{15}\int (-\tau_{0j}) Y_{2m,j}^{B*}r^2 d^{3}x ,
\end{eqnarray}
where $\tau$ is the stress-energy tensor, $Y_{lm}$ are scalar and $Y^{B}_{lm}$ magnetic vector spherical harmonics.

In a glitching pulsar a quadrupole can arise essentially in one of two ways: either by the creation of a static, asymmetric deformation, a so-called `mountain', that leads to a mass quadrupole that is swept around by rotation, or by the excitation of oscillation modes of the star (see \citealt{HasSchwenzer} for a review), which can lead to both mass and current contributions to the quadrupole.

\subsection{Modes}

The modes of oscillation that are the most relevant for our discussion are the $f$-mode (or fundamental mode) and the $r$-mode.
The $f$-modes are polar in nature, and emit GWs mainly via the mass quadrupole; see e.g.\ \citet{ks_99}. They are `shape' deformations of the star, as the perturbation due to the mode has no nodes, and the frequency of the mode $\omega$ scales with the average density of the star, i.e. $\omega\propto \sqrt{M/R^3}$, with $M$ the mass and $R$ the radius of the neutron star. The $f$-mode is an efficient GW emitter, and this leads also to rapid damping, on a timescale (for the lowest order mode, with $l=2$) $\tau\propto \omega^{-6} R$, which for typical NS parameters is less than a second. The mode will thus be rapidly damped. Note that if one can measure both the frequency and damping timescale, the different scalings would, in principle, allow to measure mass and radius of the star, and thus constrain the equation of state; see e.g.\ \citet{ak_98}.

The $r$-modes, on the other hand, are axial modes, that are only present in rotating stars, and emit GWs mainly via the current quadrupole, as to leading order they do not couple to density and pressure perturbations in the star. The velocity perturbation due to an $r$-mode can be written as: \citep{Lindblom:1998wf}
\begin{equation}
\delta\vec{v}=\alpha R\Omega\left(\frac{r}{R}\right)^{m}\vec{Y}_{mm}^{B}\!\left(\theta,\varphi\right)\mathrm{e}^{i\omega t} , \label{eq:r-mode}
\end{equation}
where $R$ is the radius of the source and $\alpha$ is a dimensionless amplitude parameter. Furthermore the mode frequency $\omega$ is linked to the angular  velocity of the star $\Omega$ by the simple relation $\omega =4/3 \Omega$, although note that there will be corrections to this relation due to rapid rotation and general relativity.

The $r$-mode is particularly interesting as it can be driven unstable due to GW emission, via the so-called Chandrasekhar-Friedman-Schutz (CFS) mechanism, even at low rotation rates \citep{NilsReview}, and may be active also in some younger glitching pulsars, such as PSR J0537-6910 \citep{Alford14}. Note that in principle the $f$-mode can also be CFS unstable, but in this case the instability only sets in at much higher rotation rates, and the mode is expected to be damped rapidly by superfluid mutual friction in the general population of glitching pulsars.

\subsection{Mountains}

To describe the detectability of the signals in the various scenarios, one often introduces an intrinsic strain $h_0$; see e.g.\ \citet{Riles23}. This is the amplitude that would be measured in the hypothetical case of a detector positioned at one of the Earth's poles and the source vertically above it, with its rotation axis parallel to that of the Earth.
For GWs due to deformations of the star, the main contribution to the intrinsic strain amplitude $h_0$ is due to the mass quadrupole, and takes the form:
%\cite{Owen:2009tj}
%\begin{equation}
%h_{0}=4\pi^2 G \frac{I_{zz} f^2 \epsilon}{D}
%\end{equation}
\begin{equation}
    h_{0}=\sum_{m=-2,2} 4\pi^2 \frac{G}{c^4}\sqrt{\frac{5}{8\pi}}\frac{f^2}{D}|I_{2m}| ,
\end{equation}
where $D$ is the distance to the source, and $f$ the GW frequency. The emission will be at the spin frequency $\nu=\Omega/2\pi$, i.e. $f=\nu$ for the $m=-1$ and $m=1$ components, and at $f=2 \nu$ for the $m=2$ component, which is expected to provide the dominant contribution for most deformation mechanisms (see \citealt{Riles23} for a review). In fact many observational GW papers use a slightly different definition of the mass quadrupole $Q_{22}=\int \Re{(\delta\rho_{22})} r^2 dr$, where $\delta \rho_{22}$ is the $l=m=2$ component of the density perturbation, such that
\begin{equation}
    h_{0}=4\pi^2 \frac{G}{c^4}\sqrt{\frac{8\pi}{15}}\frac{f^2}{D} Q_{22} .
\end{equation}.

This definition is often presented in terms of an ellipticity $\epsilon$, defined as:
\begin{equation}
 \epsilon=\frac{I_{xx}-I_{yy}}{I_{zz}}=\frac{Q_{22}}{I}\sqrt{\frac{8\pi}{15}}\, ,  
\label{epsilondef} \end{equation}
where $I_{ii}$ are the principal moments of inertia and $I_{xx}\approx I_{yy} \approx I_{zz}\approx I$.
In this case we have:
\begin{equation}
    h_0=4\pi^2\frac{G}{c^4}I \frac{f^2}{D}\epsilon .
\end{equation}

\section{Spin evolution in a glitch}

As already mentioned glitches are sudden jumps in frequency that disrupt the previous, `regular' spindown. 

To understand the overall timing signature of these events it is first of all useful to separate the evolution of the pulsar's rotation rate $\nu(t)$ in a secular part $\nu_{\rm sec}(t)$ and a perturbation due to the glitch (also known as a `residual'), $\Delta\nu(t)$, such that $\nu(t)=\nu_{\rm sec}(t)+\Delta\nu(t)$. 
In the most simple model one has that, for times $t>t_g$, with $t_g$ the inferred time of the glitch
\begin{equation}
    \Delta\nu(t)=\Delta\nu_p+\Delta\dot{\nu}_p\Delta t+\Delta\nu_t\exp{-t/\tau} ,
    \label{timing}
\end{equation}
where $\Delta t=t-t_g$, a subscript $p$ indicates any `permanent' changes in frequency $\nu$ and frequency derivative $\dot{\nu}$, i.e. changes that do not decay away before the next glitch, while the subscript $t$ denotes the transient part that decays on a timescale $\tau$. Note that for simplicity we assume a single exponentially decaying term, which is sufficient for most glitches in which a relaxation is observed. For the Vela and a number of other pulsars, however, additional exponentials are necessary to fit the data, with timescales varying from minutes to months \citep{AntoReview}.

In practice we see that the timing model above naturally introduces two timescales: a very short timescale associated with the rise of the glitch (assumed in equation (\ref{timing}) to be instantaneous and thus associated with the permanent steps), and a longer timescale $\tau$ associated with the relaxation.

This division is, of course, somewhat arbitrary, but we will use it to describe the different models for GW emission in the following. The best current upper limits for the rise time suggest it is over in less than a minute \citep{Velarise1, Velarise2}, as are any rapidly decaying components of the relaxation. We will consider any dynamics occurring on timescales shorter than a minute (e.g. the excitation and damping of $f$-modes) as associated with the rise, and will describe separately longer lived signals associated with the longer timescales of days to months, associated with the relaxation.

\section{GW emission during the rise}
\label{rise}
Let us first of all focus on the initial spin-up phase of the glitch. Most models for these events rely on the vortex un-pinning paradigm that we have outlined in the introduction, i.e. a large number of vortices unpinning and moving out, leading to a rapid transfer of angular momentum. In particular quantum mechanical simulations, in which one solves the Gross-Pitaevskii equation for a superfluid in a spinning-down trap \citep{GPWarszawski}, have shown that vortices can knock each other on as they move out, leading to large avalanches that can, at least for a number of pulsars, explain the size and waiting time statistics that we observe \citep{Avalanche1, Warszawski13, Avalanche2, Fuentes19}. 

The simulations also show that the non-axisymmetric vortex re-arrangement during a glitch can lead to a transient current quadrupole and a burst of GWs. The gravitational wave strain depends strongly on the unpinning geometry and the stellar parameters. In the case in which the vortex travel time is much shorter than the pulsars rotation period, as we expect in most cases, vortices move mostly azimuthally, and one can estimate that \citep{Warszawski12}:
\begin{equation}
h_0\approx 10^{-28} \left(\frac{D}{1\mbox{kpc}}\right)^{-1}\left(\frac{\tau}{1\mbox{ms}}\right)^{-2}\left(\frac{\Delta\Omega/\Omega}{10^{-7}}\right)\left(\frac{\Omega}{10^{2}\mbox{rad/s}}\right) \label{riseh1} ,
\end{equation}
where $\tau$ is the glitch rise time. If, on the other hand, vortices can travel a significant radial distance during a glitch, one has

\begin{equation}
h_0\approx 10^{-24} \left(\frac{D}{1\mbox{kpc}}\right)^{-1}\left(\frac{\Delta r}{1\mbox{cm}}\right)^{-2}\left(\frac{\Delta\Omega/\Omega}{10^{-7}}\right)\left(\frac{\Omega}{10^{2}\mbox{rad/s}}\right)^3 .
\end{equation}

We caution the reader that the above estimates are highly uncertain, given that the computational cost of the simulations is such that only a small number of vortices (of the order of $10^2$) can be simulated, so that scaling the results up to a realistic number of vortices in a glitch (of the order of $10^{12}$) is very challenging. Nevertheless, taking the expression in (\ref{riseh1}) at face value, the non-detection of a GW signal from the Vela glitch of 2006, can be used to set a lower limit of $\tau>10^{-4}$ ms on the glitch rise \citep{Warszawski12}.

One of the first mechanisms that was considered in this context was the excitation of the fundamental mode, or $f$-mode, following a glitch. As we have discussed previously, this mode is of particular interest as not only is it expected to be efficiently excited, but its frequency is also a direct probe of the NS average density, therefore of its mass, radius and equation of state \citep{ak_98}. By performing numerical time-evolutions of a multi-fluid star, \citet{Sideryglitch} examined this possibility, and found that following a large glitch ($\Delta\Omega/\Omega=10^{-6}$) the $l=2$ $f$-mode is excited, but the associated GW amplitude is generally too weak to be detectable even by third generation detectors, although the effect of stratification in the models can make a very large difference in the estimates.

The problem was re-analysed recently by \citet{HoFmode} who considered GW emission due to transient $f$-modes excited in the known glitching pulsar population, using realistic equations of state to calculate the mode frequency and damping timescale. They estimate a peak amplitude of:
\begin{eqnarray}
    h_0&=&7.21\times 10^{-24}\left(\frac{1\mbox{kpc}}{D}\right)\left(\frac{\nu_s}{10\mbox{ Hz}}\right)^{1/2}\left(\frac{\Delta\nu_s}{10^{-7}\mbox{ Hz}}\right)^{1/2}\nonumber\\
&&\left(\frac{1\mbox{kHz}}{\nu_{gw}}\right)\left(\frac{0.1\mbox{s}}{\tau_{gw}}\right) ,
\end{eqnarray}
where $\nu_s$ is the spin frequency of the star, $\Delta\nu_s$ the jump in frequency associated with the glitch and $\nu_{gw}$ and $\tau_{gw}$ the mode frequency and damping timescale (which correspond to the frequency and decay time of the GW signal).
This estimate suggests that the largest glitches observed in the pulsar population may excite $f$-modes that could be detected by next generation observatories. This would not only allow for constraints on the stellar mass and radius (and thus the EoS), but also for the exciting prospect of detecting GW signals also from pulsars that undergo glitches that are not electromagnetically observed.

Another possibility is that the glitch is triggered by a fracture in the solid crust.  Such a starquake could directly excite f-mode oscillations, and overtones of the f-mode, i.e. oscillations with multiple modes in their radial eigenfunctions.   A simple model of the spectrum of such modes excitations was explored by \citet{KJ_15}, who looked at the simple case where the crust in a spinning-down star suddenly cracks everywhere, all at once, assuming a new zero-strain configuration.  \citet{KJ_15} found that the dominant excitation was of a mode of similar character to the f-mode, with an amplitude of order $\delta r / R \sim 10^{-6}$, where $\delta r$ is the typical size of the displacement of a fluid element due to the glitch.  Excitations of this size are too small to be detected by current gravitational wave detectors, so, within the simple \citet{KJ_15} model at least, the prospects for direct detection are not good.

In addition to the standard glitches considered above, pulsar astronomers have identified a numbers of small so-called ``glitch candidate'' events, in both the Crab \citep{espinoza_etal_14} and Vela \citep{espinoza_etal_21} pulsars.  These are smaller in magnitude than standard glitches, and can come in both conventional spin-up ($\Delta \nu >0$) and spin-down ($\Delta \nu < 0$) forms.  Their relationship to conventional glitches remains unclear.  Motivated by the dual sign of $\Delta \nu$, \citet{YimJones23fmode} suggested that these might represent small non-axisymmetric starquakes.  By conservation of angular momentum, quakes that excite prograde modes would have $\Delta \nu <0$, while retro-grade modes would have $\Delta \nu > 0$.   This mode excitation and subsequent rapid decay leads to a burst of GWs, and angular momentum is radiated to infinity, leading to the visible jump in frequency \citep{YimJones23fmode}. 

Assuming the main contribution to be due to the $l=m=2$ component of the mode, the mode excitation amplitude $\alpha$ can be linked to the glitch amplitude by
\begin{equation}
\alpha^2=\frac{4\pi}{15}\frac{\Delta\Omega}{\Omega}\frac{\Omega}{\omega_2} ,
\end{equation}    
where $\omega_2=\sqrt{4GM/5R^3}$ is the $f$-mode frequency.  The GW signal associated to the rise will decay on the $f$-mode damping timescale $\tau=10c^5/\omega_2^6 R^5$ and takes the form:
\begin{equation}
    h_0=\frac{4}{25}\sqrt{\frac{30}{\pi}}\alpha\frac{(GM)^2}{c^4R}\frac{1}{D}\exp{-t/\tau} .
\end{equation}
\citet{YimJones23fmode} inserted numbers for known glitching pulsars, and found that, for the Vela pulsar, one obtains $\alpha \sim 10^{-6}$, and that the model is directly testable when third generation GW detectors go online.  Even more speculatively, one could imagine applying this model to glitches themselves (rather than glitch candidate events).  This would require rather large ($\alpha\gtrsim 10^{-4}$) mode excitations, but would be correspondingly easier to detect.

\section{GW emission during the relaxation }

Another promising possibility is that, following the glitch itself, independently of the mechanism causing the rise, modes of oscillation may be excited, or a mountain created. Rather than a short burst of gravitational radiation, as in the case of emission associated with the rise we considered earlier, this would lead to a `transient' continuous signal, which decays on the typical timescales of the glitch relaxation, from hours to months.  Several methods have been developed to search for such signals, that are somewhat in between bursts and traditional longer lived continuous waves, in GW data (see e.g. \citet{Prix2011}), and searches have been carried out on glitching pulsars, as we will see in the following.

While the (very short-lived) $f$-modes described previously are expected to be more efficient GW emitters, the possibility of exciting longer lived $r$-modes after the glitch rise was considered by \citet{Santiago-prieto}, who however found that in this case the signal is weaker, and is unlikely to be detected even by third generation detectors. The situation is different if the $r$-mode is not simply excited by the glitch, then damped by viscosity, but can grow unstable, leading to large amplitude GW emission between glitches. We will consider this scenario in the next section.

Another interesting possibility is that part of the energy released during the glitch may create a transient mass quadrupole, or `mountain' on the star \citep{Prix2011}. If the mountain is large enough to lead to a GW torque that is comparable to the electromagnetic torque, it may also explain the increased spin-down rate $|\dot{\nu}_s|$ which is generally observed after a glitch.  

%It is first of all useful to separate the evolution of the pulsar's rotation rate $\nu{t}$ in a secular part $\nu_{sec}(t)$ and a perturbation due to the glitch (also known as a `residual'), $\Delta\nu(t)$, such that $\nu(t)=\nu_{sec}(t)+\Delta\nu(t)$. In the most simple model one has that, for times $t>t_g$, with $t_g$ the time of the glitch
%\begin{equation}
%    \Delta\nu(t)=\Delta\nu_p+\Delta\dot{\nu}_p\Delta t+\Delta\nu_t\exp{-t/\tau}
%\end{equation}
%where $\Delta t=t-t_g$, a subscript $p$ indicates any `permanent' changes in frequency $\nu$ and frequency derivative $\dot{\nu}$, that do not decay away before the next glitch, while the subscript $t$ denotes the transient part that decays on a timescale $\tau$. Note that for simplicity we assume a single exponentially decaying term, which is sufficient for most glitches in which a relaxation is observed. For the Vela and a number of other pulsars, however, additional exponentials are necessary to fit the data, with timescales varying from minutes to months.

Starting from this assumption \citet{2020MNRAS.498.3138Y} were able to use EM observations of pulsar glitches to estimate the strength of the GW signal that can be expected in such a model.  Using the assumption that the increase in spindown rate is due to GW torques from a transient mountain (and assuming that no permanent mountain remains on the NS, therefore no permanent component of the spindown rate is present in the relaxation. Note that this is true in most pulsars, but, for example, not in the Crab), the initial ellipticity can be written as:
\begin{equation}
    \varepsilon=\sqrt{-\frac{5}{32(2\pi)^4}\frac{c^5}{G I}\frac{\Delta\dot{\nu}_t}{\nu^5}} , \label{mountain}
\end{equation}
where $I$ is the moment of inertia of the star, and note that, as the spindown rate increases, $\Delta\dot{\nu}_t$ is negative. As $\Delta\dot{\nu}_t$ decays as $\exp{-t/\tau}$, from equation (\ref{mountain}) we see that the ellipticity will decay on a timescale $\tau_{GW}=2\tau$, so that the corresponding GW strain will be:
\begin{equation}
   h_0=\sqrt{-\frac{5}{2}\frac{G}{c^3}\frac{1}{D^2}\frac{\Delta\dot{\nu}_t}{\nu}} \exp{-t/2\tau} .
\end{equation}

\citet{2020MNRAS.498.3138Y} calculated the expected strain for the currently known population of glitching pulsars and found that in general current detectors would struggle to detect these signals from most pulsars, except for the Vela pulsar, which is a promising target for current and future observational runs of the LVK network. Next generation detectors such as the Einstein Telescope (ET; \citet{ET_2020}) or Cosmic Explorer (CE; \citet{CE_2020}), on the other hand, should be able to confidently detect these transient signals from most glitching pulsars, should this emission mechanism be active.

Following the glitch rise it is also the case that the interior fluid must 'catch' up, leading to non-axisymmetric Ekman flows in the star. \citet{Bennett} estimated the current quadrupole that may result from such a flow, and while their model did not include the effect of the magnetic field on the coupling timescales (although see \citet{MelatosFossil} for a discussion of the effect of magnetic fields on Ekman flows), and relied on calculations in cylindrical geometry, they found that the characteristic wavestrain depends on the EoS, but scales roughly as
\begin{equation}
    h_0\approx 10^{-27}\left(\frac{\nu}{10 \mbox{Hz}}\right)^3\left(\frac{1 \mbox{kpc}}{D}\right)\left(\frac{\Delta\nu/\nu}{10^{-6}}\right)\left(\frac{I_c/I_s}{10^{-2}}\right)\left(\frac{\Delta r/R}{10^{-6}}\right)^{-1} ,
\end{equation}
where $I_c$ is the moment of inertia of the crust, $I_s$ the moment of inertia of the superfluid and $\Delta r$ the radial distance travelled by the vortices during the glitch. The results were recently updated by \citet{singh17} to include a wider range of parameter space for the nuclear equation of state, and the mass quadrupole contribution.

These results suggest that, despite the fact that only a small fraction of the glitch energy will be radiated in GWs by this mechanism (of the order of a part in $10^7$) current searches may be able to detect such long duration signals following the largest glitches, should one be fortunate enough to have such an event occur during LVK observations. Next generation detectors, on the other hand, should be able to probe these models also for smaller glitches, extending the searches to a larger portion of the known (and unknown) glitching pulsar population, and possibly putting constraints on the compressibility of nuclear matter in NS \citep{VE1, VE2, Bennett}.

{\bf Another mechanism that could produce transient gravitational waves following a glitch was proposed very recently by \citet{yim_etal_23}.  Their model was motivated in large part by the glitches and \emph{anti-glitches} seen in magnetars, but could equally well be applied to glitches in the main pulsar population.  The new idea was that the glitch (or anti-glitch) is caused by the sudden ejection of material from the star's surface, with this material then becoming trapped in the magnetosphere.  Depending upon the exact geometry, and on the star's equation of state, this can produce either a sudden increase or sudden decrease in the star's moment of inertia, manifesting as a glitch or anti-glitch, respectively.  This model naturally produces a shift in the orientation of the principle axes of the star's moment of inertia tensor.  The star is therefore set into free precession, emitting transient GWs at once and twice the spin frequency (see e.g. \citep{zs_79, ja_02}).  This dual-harmonic emission could be allowed for in a targeted GW search.}

\section{Long-lived GW emission between glitches}

While the mechanisms we have discussed above are associated with the glitch relaxation, and thus decay on the typical timescales of hours-months associated with this process, it is also possible that non-asymmetries created by the vortex pinning and unpinning process may persist, that modes of oscillation may grow unstable, or that permanent mountains may be formed, leading to substantial GW emission over secular timescales.

With regard to the superfluid, current and mass quadrupoles due to vortex rearrangement in past glitches may remain frozen in \citep{MelatosPersist}, and large mass quadrupoles may also result from vortex pinning to superconducting flux tubes in the core of the star.  Recent analysis of the rapid relaxation \citep{HaskellCrustCore} and of the activity of the Vela pulsar \citep{CrustNotEnough, ChamelCrust} has, in fact, suggested that part of the core moment of inertia should be involved in the glitch. This is possible if part of the pinned superfluid is in the outer core, where neutron vortices may pin to superconducting fluxtubes in the strong toroidal field regions \citep{AlparCore1, AlparCore2} (although note that equilibrium models of magnetic fields in superconducting stars show that the toroidal field can be entirely expelled to the crust of the star, see e.g. \citealt{LanderSupercon, AnkanSupercon}). If this is the case vortex accumulation can lead to asymmetric flows and, if the rotation axis is inclined with respect to the magnetic axis, generate mass quadrupoles \citep{HaskellPiUltimo}. Such quadrupoles may explain the apparent residual ellipticity of the millisecond radio pulsars, which has been suggested as an explanation for the lack of systems with low values of the period $P$ and period derivative $\dot{P}$ observed in this population \citep{Woan}. In this scenario, such systems cannot exist as they are spun out of the aforementioned region of parameter space by GW torques, and the associated signals may be detectable by next generation observatories such as ET and CE. 

A particularly interesting glitching pulsar is  PSR J0537-6910. This is a young X-ray pulsar, and also the most frequent glitcher, as it undergoes large glitches roughly every 100 days. Furthermore it is the only pulsar for which there is a strong correlation between the size of a glitch and the waiting time until the next, therefore allowing us to predict when the next glitch will occur \citep{Middleditch, Ant0537}. A detailed analysis of the spin evolution during the post-glitch relaxation has also revealed that the braking index $n=\ddot{\Omega}{\Omega}/{\dot{\Omega}}$ tends asymptotically towards $n=7$ \citep{2018ApJ...864..137A, Ferdman0537, Ho0537}. This is an interesting results, as one would expect $n\approx 3$ if the main contribution to the spin-down torque is due to the emission of EM waves. If, on the other hand, GW torques dominate the spin-down, one expects $n=5$ for mountains, and $n=7$ for $r$-modes. Furthermore, it has been suggested that young, hot and rapidly rotating pulsars are spun-down to the periods observed in the standard pulsar population by GW emission due to unstable $r$-modes early in their life, and a theoretical analysis by  \citet{Alford14} point exactly at PSR J0537-6910 as a system that may still be young enough to be at the end of its $r$-mode driven evolution. 

Searches for this r-mode emission  have been carried out in GW data from the O1, O2 and O3 runs of the LVK network \citep{Fesik, LVK0537rmode}, combined with X-ray timing with NICER \citep{Ho0537}, which is necessary to track the spin-evolution of the system, thus allowing to search around the glitch epochs. The most recent searches \citep{LVK0537rmode} have significantly restricted the parameter space allowed for this scenario, putting limits on the mass of the star and its EoS. Increased sensitivity in the next observational runs may thus allow us to detect such a signal, constraining the EoS of dense matter, or exclude this scenario, therefore pointing towards internal torques to explain a braking index of $n=7$.

Alternatively, one can consider the possibility that a glitch may be accompanied by the formation of a mountain.  This was suggested in the context of spinning-up (i.e. accreting) stars by \citet{fhl_18}, who noted that the spin-up process necessarily produces a centrifugal deformation, that strains the crust.  The centrifugal force itself is axisymmetric.  However, it is perfectly plausible that the cracking event that relieves the stress should occur in a non-axisymmetric way.  This might  be the case for an event that started at one particular point, and then propagated outwards.  As long as the crust is spun-up enough, and the crustal breaking strain is low enough, than a sufficiently rapidly spin-up start might gain a mountain through just such a process.

This mainly qualitative argument was examined in more detail by \citet{gc_22}, who argued that the triaxial shape of the star immediately after the glitch must be bounded by two axisymmetric configurations, one corresponding to the strained  star just before the starquake, and the other (more oblate one) to the shape the star would have if all strain were relieved at fixed angular momentum.  In this way, they found that the maximum mountain size should lie in the range $\epsilon \sim 10^{-9}$--$10^{-5}$, depending upon the mass and equation of state of the star.

{This problem was examined also by \citet{Kerin22}, who studied, with the aid of a cellular automaton model, the ellipticity that can be built up as a rapidly rotating neutron star spins-down, and the strain relaxes due to crust quakes. They found that if the star is born rotating faster than $\nu\approx 750\,$ Hz, tectonic activity will continue until it has spun down to approximately $1\%$ of its initial frequency, and lead to an average ellipticity between $10^{-13}\lesssim \epsilon \lesssim 10^{-12}$ , which however can have large excursions between events. Such a value would lead to long lived emission that is not strong enough for current interferometers, but may be detectable by future gravitational wave observatories}.

Note that these arguments do not take into account that in addition to spinning-up the star, the accretion process replaces the crust.  Presumably, the newly formed crust would not inherit the full  strain of the pre-accreted crust.  The extend to which this would reduce the maximum mountain that can be built through such a cracking process is not clear, and worth investigating.

\section{Searches}

A number of searches have been carried out for GW signals associated with pulsar glitches, some looking for short bursts, others looking for longer signals linked to the post-glitch relaxation.  For the longer duration signal, as first discussed in \citet{Prix2011}, one has to search for a `transient' continuous signal, i.e. a signal $h$ that at time $t$ looks like a persistent continuous gravitational wave $h(t,\mathcal{A},\lambda)$, with amplitude parameters $\mathcal{A}$ and phase parameters $\lambda$, additionally modulated by a window function $\omega (t,t_0,\tau)$, which effectively
limits the duration of the CW (with $t_0$ the time of the glitch, and $\tau$ a timescale that defines the duration of the signal), so that one can write:
\begin{equation}
h(t,\mathcal{A}, \lambda, \mathcal{T})=\omega\,(t,t_0,\tau) h(t,\mathcal{A}, \lambda)\, ,
\end{equation}
where $\mathcal{T}$ represents the transient parameters, i.e. the shape of the window function $\omega$, the time of the glitch $t_0$ and timescale $\tau$. For example in the simplest case the window function will just be a damped exponential with a timescale $\tau$ associated with a relaxation timescale. A more detailed description of the search method, and discussion of the application to searches in O3 data is given in \citet{Mod21}.

One of the first such searches was a search for short duration signals associated with $f$-modes following the large Vela glitch of 2006 \citep{LSC2006glitch}, which occurred during the 5th science run of the detectors (S5), using the method developed by \citep{Clark07}. This was followed by a search for long-duration transients associated with the relaxation (timescale from hours up to
120 days) made on the O2 open data, with both Vela and Crab having glitched during that time \citep{keitel19}. In both cases no detection was made and upper limits were set on the gravitational wave amplitude following the glitch. {\bf In particular, for the 2016 Vela glitch that occurred during this period, the upper limits were close to the indirect upper limit obtained by assuming all the energy of the glitch is radiated in GWs.}

In O3 data searches were carried out for long duration transients (hours to months) from known pulsars \citep{LVKtransient}, also in the aftermath of glitches for 6 pulsars that had glitched during the observing run (including PSR J0537-6910 that glitched 4 times). No signal was detected, and in all cases the upper limits derived are above the theoretical upper limit obtained by simply assuming that all the energy released by the glitch is radiated in GWs \citep{Prix2011}. For several of the targets however the observation upper limits are not far off from the theoretical ones, and future advances in sensitivity, also using for example machine-learning based methods \citep{Mod23} may soon allow for observational constraints on the glitch mechanisms from GW searches in the next observational runs of the detectors.

In fact, \citep{Lopez22fmode} have shown that gravitational wave searches for $f$-modes following glitches could be used to detect glitches that are not observed electromagnetically, and find that in the next (fifth) observing run of the LVK network, the minimum detectable glitch size will be of $\Delta\nu\approx {10}^{-5}$  Hz, for pulsars with spin frequencies and distances comparable to the Vela pulsar.

\subsection{Stochastic Gravitational Wave Background due to glitches}

Searches have also been carried out, in data from the first three observational runs of the detectors, for a Stochastic Gravitational Wave Background (SGWB) due to the overlap of many burst-like signals \citep{DeLillo_search}, assumed to be due to avalanche dynamics, as proposed by \citet{Warszawski12} and described in section \ref{rise}. The idea behind this is that there is indeed a large number of electromagnetically dark neutron stars in the galaxy, that glitch roughly at the rate suggested by the observed glitching pulsar population, an overlap of GW bursts during the rise may lead to a detectable signal.
No signal was detected, but \citet{DeLillo_search} demonstrated that the method can be used to constrain parameters of vortex avalanche models. In particular, if current theoretical and observational efforts are successful in determining the expected glitch rate, this can break degeneracies in the models, and allow to use such searches for a SGWB to constrain glitch rise times and the distance vortices move in the crust.

\section{Conclusions}

As we hope we have made clear, the potential association between pulsar glitches and gravitational wave astronomy opens a new door to probe the glitch mechanism itself.  There is a rich set of physical mechanisms at play, where glitch-induced gravitational waves could last from anything between fractions of a second to more or less indefinitely.  The behaviour of the elastic crust, the  neutron  superfluid, and the proton superconductor, are all relevant.  The observational challenges of detecting such GW signals are severe, but the potential pay-off, both for pulsar physics and for probing the high density equation of state, is very large. Future joint electromagnetic and gravitational observations of signals from glitching pulsars may, therefore, play a key part in developing our understanding of high density physics in neutron star interiors.

\section*{Acknowledgements}

DIJ acknowledges support from the STFC via grant number ST/R00045X/1. BH acknowledges support from the National Science Center Poland (NCN) via OPUS grant 2019/33/B/ST9/00942.

\appendix

\bibliographystyle{elsarticle-harv} 
\bibliography{main}



%% For example:

%% \bibitem[Aladro et al.(2015)]{Aladro15} Aladro, R., Martín, S., Riquelme, D., et al. 2015, \aas, 579, A101

%%\end{thebibliography}

\end{document}